%% file: feshbach_paper_arxiv.tex
\documentclass[aps,pra,superscriptaddress,reprint,floatfix,longbibliography,nofootinbib]{revtex4-2}

\usepackage{multirow}
\usepackage[percent]{overpic}
\usepackage{etoolbox}
\apptocmd{\thebibliography}{\raggedright}{}{}
\usepackage[section]{placeins}

\usepackage{array}
\usepackage{amsmath,amssymb}
\usepackage{graphicx}
\usepackage{soul}
\usepackage{color}
\usepackage{physics}   % tell physics not to define \qty
\usepackage{booktabs}
\usepackage{siunitx}
\usepackage{array}
\usepackage[normalem]{ulem}
\usepackage{adjustbox}
\usepackage{bm}
\usepackage{textcomp} % This package is just to give the text quote '
\usepackage{enumitem}
\usepackage{hyperref}
\usepackage{tabularx}
\usepackage{tikz}
\usepackage{bbold}
\usepackage{pgfplots}
\usepackage{amsfonts}
\AtBeginDocument{\RenewCommandCopy\qty\SI}

\usepackage{CJKutf8}
\usetikzlibrary{decorations.markings,arrows}
\usetikzlibrary{shapes.geometric}
\hypersetup{
colorlinks=true,
urlcolor= blue,
citecolor=blue,
linkcolor= blue,
}
\usepackage{wasysym} % for "\varhexagon" macro

\renewcommand{\vec}[1]{\bm{#1}}

\definecolor{amethyst}{rgb}{0.6, 0.4, 0.8}
\definecolor{blue-violet}{rgb}{0.54, 0.17, 0.89}
\pgfplotsset{compat=1.18}

\newcommand{\MB}{\mathrm{MB}}
\newcommand{\AZB}{\mathrm{AZB}}
\newcommand{\TDB}{\mathrm{TDB}}

\begin{document}

\begin{CJK}{UTF8}{gkai}

\title{Characterization of Feshbach resonances in a $^6\mathrm{Li}{-}^7\mathrm{Li}$ mixture using improved interaction potentials}

\author{Jing-Chen Zhang(张京晨)}
\email{jczhang@terpmail.umd.edu}
\affiliation{Institute for Physical Science and Technology, University of Maryland, College Park, Maryland 20742, USA}
\affiliation{Department of Chemistry and Chemical Biology, University of Maryland, College Park, Maryland 20742, USA}
\affiliation{Joint Quantum Institute, University of Maryland and National Institute of Standards and Technology (NIST), College Park, Maryland 20742, USA}

\author{Paul Julienne}
\affiliation{Joint Quantum Institute, University of Maryland and National Institute of Standards and Technology (NIST), College Park, Maryland 20742, USA}

\author{Yu Liu}
\email{yuliu@umd.edu}
\affiliation{Institute for Physical Science and Technology, University of Maryland, College Park, Maryland 20742, USA}
\affiliation{Department of Chemistry and Chemical Biology, University of Maryland, College Park, Maryland 20742, USA}
\affiliation{Joint Quantum Institute, University of Maryland and National Institute of Standards and Technology (NIST), College Park, Maryland 20742, USA}

\date{\today}

\begin{abstract}
We characterize Feshbach resonances in all isotopologs of the $\mathrm{Li}{-}\mathrm{Li}$ system with improved interaction potentials. Starting from spectroscopically accurate Morse/long-range (MLR) potential-energy curves for the singlet ($X^{1}\Sigma^{+}$) and triplet ($a^{3}\Sigma^{+}$) electronic states of $\mathrm{Li}_2$, we apply small phenomenological inner-wall adjustments [following \href{https://doi.org/10.1103/PhysRevA.89.052715}{Julienne and Hutson, Phys.\ Rev.\ A 89, 052715 (2014)}] and fit the resulting potentials to threshold measurements for the $^6\mathrm{Li}{-}^6\mathrm{Li}$ and $^7\mathrm{Li}{-}^7\mathrm{Li}$ isotopologs, including binding energies, scattering lengths, and Feshbach resonance positions. Using the optimized potentials in coupled-channel scattering calculations, we predict and characterize s-wave Feshbach resonances in the $^6\mathrm{Li}{-}^7\mathrm{Li}$ isotopolog. In its lowest-energy hyperfine channel, all resonances are narrow ($\sim 0.01$--$0.1$~G), strongly closed-channel dominated, and predominantly triplet in electronic spin character, in marked contrast to the homonuclear systems. These results provide a foundation for designing Raman optical-transfer pathways to produce ultracold $\mathrm{Li}_2$ molecules in deeply bound rovibrational levels of both the $X^1 \Sigma^{+}$ and $a^3 \Sigma^{+}$ potentials across all three isotopologs.

\end{abstract}
\maketitle
\end{CJK}
\section{Introduction}

Scattering resonances are a quantum effect that occurs in collisions~\cite{Mott1965}. At resonance, a pair of colliding particles is temporarily trapped in a weakly bound collision complex. This occurs when the two-body scattering state strongly couples to either a quasi-bound state belonging to the same molecular potential (a shape resonance) or a bound state belonging to a different potential (a Feshbach resonance). Experimentally, collisional systems can be brought into resonance by tuning the energy of participating scattering and bound or quasi-bound states to be degenerate\cite{Chin2010Rev.Mod.Phys.}. This can be accomplished by tuning the kinetic energy of the scattering state, the energy of the collision threshold, and/or the energy of the bound or quasi-bound state. The location of a resonance in energy depends on the shape of the molecular potential underlying the collision, while its width can be minuscule compared to the energy scale of the potential. These features make scattering resonances a highly sensitive probe of molecular interactions.

Magnetically tunable Feshbach resonances play an important role in ultracold atomic and molecular physics \cite{Chin2010Rev.Mod.Phys.}. These resonances leverage the fact that the scattering state (open channel) and the bound state (closed channel) can have different magnetic dipole moments and thus experience differential Zeeman shifts. By tuning the magnitude of an applied magnetic field ($B$), these channels can be brought in and out of resonance. For the $s-$wave dominant collisions prevalent at ultralow temperatures ($T\lesssim1\mu$K), the drastic change in scattering length ($a$) around a Feshbach resonance allows for tuning of the interparticle interaction over a wide dynamical range. This control has enabled ultracold atoms to become a versatile platform for quantum technologies—especially quantum simulation\cite{Georgescu2014Rev.Mod.Phys.,Cornish2024NaturePhysics}.

Feshbach resonances also play a vital role in the production of ultracold diatomic molecules \cite{Ni2009Phys.Chem.Chem.Phys.}. In the widely used coherent association scheme, which brought us the first high phase-space density ultracold molecular gases, a pair of colliding ultracold atoms in the open channel are adiabatically converted into a weakly bound state by ramping the $B$ field across a Feshbach resonance. The resulting complex (i.e., ``Feshbach molecule'') could then be transferred to a more deeply bound molecular state using stimulated Raman adiabatic passage (STIRAP). This method has been particularly successful in forming bialkali molecules (e.g., Cs$_2$\cite{Danzl2008Science,Danzl2010NaturePhysics}, KRb \cite{Ni2008Science}, LiNa \cite{Rvachov2017Phys.Rev.Lett.}), which serve as building blocks of molecule-based quantum technology \cite{Cornish2024NaturePhysics}, as well as subjects for investigating molecular collisions and reactivity at the quantum level \cite{Liu2022AnnualReviewofPhysicalChemistry}. 

A hallmark in the study of scattering resonances is the strong synergy between theory and experiment --- experimental measurements of the predicted resonances provide benchmarks for theory, leading to refined models for molecular interaction with improved predictive power for scattering observables. In the context of ultracold atomic collisions, it has been shown that highly predictive models of threshold physics can be obtained by fitting interaction potentials with only a few parameters, which accurately capture the long-range behavior while allowing flexibility of the short-range potential\cite{Cote1994Phys.Rev.A}. Berninger \textit{et al.}\cite{Berninger2013Phys.Rev.A} fitted six parameters: two describing the long-range interaction and four describing the short-range part to obtain high-precision predictive models for Cs$_2$ bound states near the threshold. This approach was also used successfully by Julienne and Hutson to characterize Feshbach resonances and bound-state properties in the $^6\mathrm{Li}{-}^6\mathrm{Li}$ and $^7\mathrm{Li}{-}^7\mathrm{Li}$ systems\cite{Julienne2014Phys.Rev.A}.

Although effective, the fitting strategy above has limitations. Its performance often hinges on adding a \textit{shift} term to each electronic potential,
\begin{equation} 
\label{shiftTerm} 
V_{\mathrm{shift},S}(R)=\mathcal{S}_{S}\,(R-R_{e,S})^{2}, \qquad R<R_{e,S}, 
\end{equation} 
where $S$ labels the electronic spin, $\mathcal{S}_{S}$ sets the magnitude of the shift term, $R$ is the internuclear distance, and $R_{e,S}$ is the position of the minimum for that potential. This \textit{ad hoc} modification of the inner wall is intended to correct near-threshold inaccuracies of the underlying electronic potentials and thereby improve the accuracy of the predicted threshold observables. However, the reduced $\chi^{2}$ we inferred from the scattering observables in Ref.~\cite{Julienne2014Phys.Rev.A} is appreciably larger than unity, and its predicted binding energies for the last bound vibrational levels of the $X(1^{1}\Sigma_{g}^{+})$ and $a(1^{3}\Sigma_{u}^{+})$ potentials of ${}^{6}\mathrm{Li}_2$ differ from the subsequent high-precision measurements of Ref.~\cite{Semczuk2014Phys.Rev.Lett.} (data not available at the time of Ref.~\cite{Julienne2014Phys.Rev.A}) by more than the experimental uncertainties. These considerations motivate renewed efforts to construct more accurate interaction potentials for lithium dimers.

In this work, we construct a model for Li($2s$)--Li($2s$) interaction potentials that more accurately characterizes the threshold physics than previous studies\cite{Kempen2004Phys.Rev.A, Julienne2014Phys.Rev.A}. Using spectroscopically accurate Morse/long-range (MLR) analytical PECs for the $X^1\Sigma^+$ and $a^3\Sigma^+$ states modified by the corresponding shift terms [Eq.~\eqref{shiftTerm}], we refit the $^{6}$Li--$^{6}$Li and $^{7}$Li--$^{7}$Li threshold data and obtain a reduced $\chi^{2}$ of 1.41—nearly $100$ times lower than the $\chi^{2}$ we calculated from Ref.~\cite{Julienne2014Phys.Rev.A}. The accuracy of the newly fitted potentials is further validated by comparisons with more recent measurements of the near threshold bound states\cite{Semczuk2014Phys.Rev.Lett.}. 

With the improved PECs, we study the Feshbach resonances of mixed isotopolog $^{6}$Li--$^{7}$Li, which has received limited experimental or theoretical attention \cite{Kempen2004Phys.Rev.A,Zhang2005AIPConferenceProceedings}. We use coupled channel methods to solve the multichannel Schr\"{o}dinger equation and obtain the magnetic-field-dependent scattering length $a(B)$ and bound-state energies $E_b(B)$.  We use these results to identify Feshbach resonances associated with different atomic hyperfine entrance channels.
% , specifying each resonance's location ($B_0$), width ($\Delta$), and, in cases of inelastic resonances, decay rate ($\Gamma_B^{\mathrm{inel}}$).
The predicted locations $B_0$ of the four $s$-wave resonances in the lowest-energy hyperfine channel are matched to measured resonances reported in Ref. \cite{Zhang2005AIPConferenceProceedings}. Their agreement represents a clear improvement over previous theory \cite{Kempen2004Phys.Rev.A}, but the discrepancies still exceed the experiment uncertainties.

We further analyze and compare the spin characters and open- and closed-channel fractions of resonant states in the lowest hyperfine channels of the different isotopologs. We find the $^{6}$Li--$^{7}$Li resonances to be narrow($\sim$10--100~mG), strongly closed-channel dominated, and predominantly triplet in electronic spin character, in marked contrast to the established behaviors of the homonuclear resonances\cite{Chin2010Rev.Mod.Phys.,Julienne2014Phys.Rev.A}. The spin character is ultimately set by whether the last bound state resides in the singlet ($X^{1}\Sigma^{+}$) or triplet ($a^{3}\Sigma^{+}$) potential, and arises naturally from reduced-mass variations across isotopologs that share almost the same electronic potentials.

This paper is organized as follows. Section~\ref{Methods} introduces the modified MLR interaction potentials for Li–Li and the coupled-channels framework used to compute threshold observables, including scattering lengths and bound-state energies. Section~\ref{Results} describes the potential optimization procedure, presents predictions for Feshbach resonances in the $^{6}$Li--$^{7}$Li system, and analyzes the spin character and open/closed channel fractions of the Feshbach molecules of lithium isotopologs. Section~\ref{sec:Discussions} outlines the limitations of the present potentials and discusses avenues for improvement.

\section{Methods} \label{Methods}

\begin{figure}[tbp]
\includegraphics[width=1.00\columnwidth]{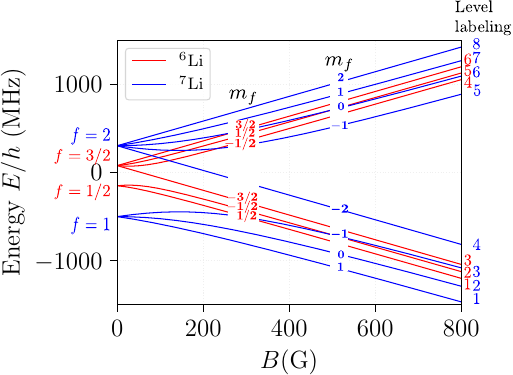}
\caption{
Zeeman shifts of the hyperfine levels in the electronic ground state ($^{2}S_{1/2}$) of ${}^{6}\mathrm{Li}$ (red) and ${}^{7}\mathrm{Li}$ (blue), labeled by the corresponding quantum numbers $f$ and $m_f$. The zero of energy is defined as the energy in the absence of hyperfine and Zeeman interactions. The numbers on the right label the atomic hyperfine states used to specify each atom in a scattering channel.
}
\label{fig: hyperfine energies}
\end{figure}

\begin{figure*}[tbp]
    \includegraphics[width=1.00\textwidth]{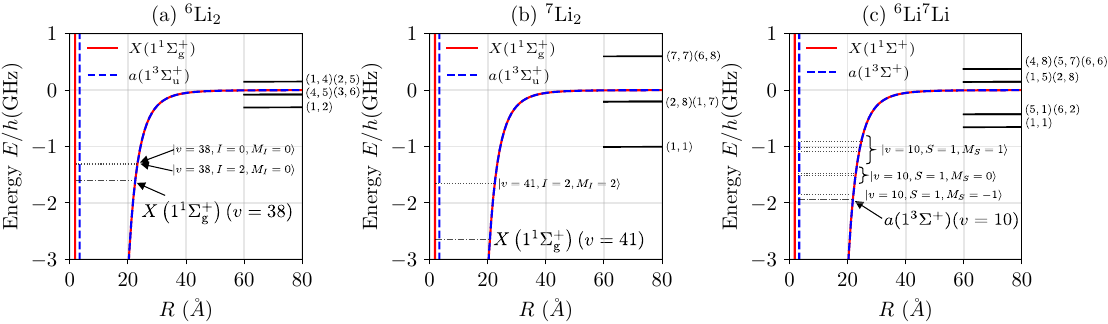}
	\caption{
    Adiabatic Born-Oppenheimer (ABO) potential energy curves for the ground singlet $X^1\Sigma^+$ and triplet $a^3\Sigma^+$ electronic states of Li$_2$ isotopologs near the dissociation threshold at $B = 0$.  Energies are shown in units of gigahertz, and $R$ represents the internuclear separation (in units of \AA). The solid horizontal lines represent energy levels of the separated atom hyperfine states which have the same $M_F$ as the lowest-energy hyperfine state (${}^{6}$Li$_2$: $M_F = 0$; ${}^{7}$Li$_2$: $M_F = 2$; ${}^{6}$Li${}^{7}$Li: $M_F = 3/2$). The dash-dot horizontal lines represent the last-bound vibrational levels supported by the ABO potentials, which do not account for hyperfine interactions. The dotted horizontal lines represent molecular bound states from the coupled-channel calculations, labeled by the approximately good quantum numbers $(v, I, M_I)$.
    }\label{fig: scattering PEC}
\end{figure*}

\subsection{Definitions and overview}
\label{sec:basis}

We first introduce various terms and concepts used in this paper to describe collisions in the ultracold regime\cite{Chin2010Rev.Mod.Phys.,Krems2009}. A collision event can be characterized as a transition between scattering states mediated by the interaction potential. The initial and final states involved are referred to as entrance and exit channels, respectively. Collisions are elastic (inelastic) if the exit and entrance channels are identical (different). At total energy $E$, channels are open (closed) if their asymptotic thresholds lie below (above) $E$, respectively, supporting extended scattering (bound) states at large interatomic separation $R$. Feshbach resonances arise when an open-channel scattering state is degenerate in energy with a closed-channel bound state, forming a quasibound resonant state.

Each scattering channel is specified by the internal quantum states of the respective particles, and the state of their relative orbital motion specified by the angular momentum quantum number $l$ and projection $m_l$. Collisions at the ultralow temperatures considered here are dominated by $l = 0, m_l = 0$ ($s$-wave) channels. For the Li($2s$) atoms considered here, the relevant internal states are the atomic hyperfine states, which can be specified in terms of the angular momentum and projection quantum numbers for the electron spin ($s, m_s$), the nuclear spin ($i, m_i$), and the coupled spin ($f, m_f$). A scattering channel can be specified in the \textit{hyperfine basis} as
\begin{equation} \label{eq:hyperfine}
    | f_a, m_{f,a}; f_b, m_{f,b}; l, m_{l} \rangle,
\end{equation}
or the \textit{totally decoupled basis} (TDB) as
\begin{equation} \label{eq:TDB}
    | m_{s,a}, m_{i,a}; m_{s,b}, m_{i,b}; l, m_{l} \rangle,
\end{equation}
which are approximately the eigenstates in the weak- and strong-field regimes, respectively. Note that the $s$ and $i$ quantum numbers are typically suppressed, since their values are known once the atomic species is specified [$s = 1/2$ for Li($2s$), $i = 1$ for $^6$Li and 3/2 for $^7$Li]. We specify the eigenstate under arbitrary magnetic field $B$ using the \textit{adiabatic Zeeman basis} (AZB), as
\begin{equation} \label{eq:AZB}
    | \alpha_a(B), m_{f,a}; \alpha_b(B), m_{f,b}; l, m_{l} \rangle,
\end{equation}
where $\alpha$ labels a Zeeman state of the atom that tends toward Eq. \ref{eq:hyperfine} in the limit of small $B$ and the limit of Eq. \ref{eq:TDB} at very large $B$. 

Figure \ref{fig: hyperfine energies} shows the energy of the atomic hyperfine states as a function of $B$ for $^6$Li and $^7$Li. Since the ordering of the hyperfine states is maintained as $B$ changes, we can designate them in order of increasing energy as 1,2,... for each species. This also simplifies the labeling of the scattering channels to a pair of such numbers. For example, the lowest-energy scattering channel for the $^{6}$Li–$^{7}$Li system is $(1,1)$, which correspond to the channel $\ket{f_{6}{=}\tfrac{1}{2},\,m_{f,6}{=}\tfrac{1}{2};\; f_{7}{=}1,\,m_{f,7}{=}1;\,\ell{=}0,m_\ell{=}0}$ in the hyperfine basis.

Bound states, on the other hand, are more naturally specified in the \textit{molecular basis} (MB)
% , with two commonly used coupling schemes. The first scheme (MB1) 
given by
\begin{equation} \label{eq:MB1}
    \bigl|S, m_S;I, m_I; \ell, m_\ell\bigr\rangle,
\end{equation}
where $S$ and $I$ are the total electronic and nuclear spins, with corresponding relations $\textbf{S} = \textbf{s}_a + \textbf{s}_b$, $\textbf{I} = \textbf{i}_a + \textbf{i}_b$, $m_S = m_{s,a} + m_{s,b}$, and $m_I = m_{i,a} + m_{i,b}$.

The total angular momentum projection $M_{\rm tot}$ is a conserved quantity:
$M_{\rm tot}
= m_{f,a}+m_{f,b}+m_\ell
= m_S+m_I+m_\ell$.  States with $S = 0$ and 1 are referred to as singlet and triplet states, respectively. For bound states, the quantum numbers $l$ and $m_l$ now specify the molecular rotation and correspond to the symbols $N$ and $m_N$ commonly used in the molecular literature\cite{Brown2003}. Our calculations and analysis require transformations between the various bases discussed here with details given in Appendix~\ref{sec:frame}. Representative transformations are depicted in Fig.~\ref{fig:transform}. The basis is properly symmetrized for identical particles in all coupled-channel calculations, as described in Appendix~\ref{sec:symmetrized}.

Figure \ref{fig: scattering PEC} depicts zero-field energies of the scattering and bound states involved in the Li--Li Feshbach resonances discussed in this work. For each system, the relevant bound states are spin sublevels of the highest vibrational state of the singlet or triplet potential: $(X^1\Sigma^+, v = 38)$ for $^{6}$Li$_{2}$, $(X^1\Sigma^+, v = 41)$ for $^{7}$Li$_{2}$, and $(a^3\Sigma^+, v = 10)$ for $^{6}$Li$^{7}$Li. As $B$ varies, the difference in the magnetic moments between the open and closed channels causes their internal energies to shift into degeneracy, leading to a Feshbach resonance [Fig.~\ref{fig:zeeman_shifts}(a)].

\subsection{Analytical descriptions of Feshbach resonances}

In the vicinity of a resonance, a system's collisional properties can differ drastically from the off-resonant case, as reflected in various scattering observables. For the isotropic \(s\)-wave collisions considered here, all effects are captured by the (generally complex) \textit{scattering phase shift} \(\delta(k,B)\), which depends explicitly on the applied magnetic field \(B\)\cite{Mott1965,Krems2009}. The phase shift is defined by the asymptotic form of the radial wave function
\begin{equation}
\psi_0(R) \xrightarrow[R\to\infty]{} \sin\!\big(kR+\delta(k,B)\big),
\end{equation}
given relative to the free particle solution \(\sin(kR)\) at the same kinetic energy \(E=\hbar^2 k^2/(2\mu)\), where $\mu$ is the reduced mass of the pair of colliding atoms.

A closely related scattering observable is the \(s\)-wave \textit{scattering length}. It is defined from the phase shift as\cite{Chin2010Rev.Mod.Phys.,Hutson2007NewJournalofPhysics}
\begin{equation} \label{eq:delta_to_a}
a(B)\equiv \lim_{k\to 0}\left[-\,\frac{\tan\delta(k,B)}{k}\right],
\end{equation}
which can be written in terms of its real and imaginary components as \( a(B)=\alpha(B)-\mathrm{i}\,\beta(B)\). Near an isolated Feshbach resonance at $B=B_0$, we classify the resonance as \textit{nondecaying} if $\beta(B)=0$ in the vicinity of $B_0$, and as \textit{decaying} if $\beta(B)>0$ due to inelastic loss. The magnetic-field dependence of the scattering length near the resonance can be captured by simple analytic forms whose parameters provide physical insights, as detailed below.

In the absence of inelastic loss channels, the scattering length is real and is described by the non-decaying resonance formula
\begin{equation}\label{eq:a_elastic}
    a(B)=a_{\mathrm{bg}}\!\left(1-\frac{\Delta}{B-B_0}\right),
\end{equation}
where $B_0$ is the field at which the resonance occurs and $a(B)$ diverges (the \textit{pole position}), $a_{\mathrm{bg}}$ is the off-resonant (\textit{background}) scattering length, and $\Delta$ is the \textit{resonance width}. In this parametrization, $\Delta$ is the separation between the pole and the zero crossing, since $a(B)=0$ at $B=B_0+\Delta$.

When inelastic decay channels are open, the scattering length becomes complex and remains finite at resonance. A commonly used decaying resonance form is\cite{Hutson2007NewJournalofPhysics,Naik2011TheEuropeanPhysicalJournalD,Nicholson2015Phys.Rev.A}
\begin{equation}\label{eq:a_inelastic}
    a(B)= a_{\mathrm{bg}} + \frac{a_{\mathrm{res}} \,(\Gamma_B^{\mathrm{inel}}/2)}{(B-B_0) + \mathrm{i}\,(\Gamma_B^{\mathrm{inel}}/2)},
\end{equation}
where the background scattering length is now complex, $a_{\mathrm{bg}}=\alpha_{\mathrm{bg}}-\mathrm{i}\beta_{\mathrm{bg}}$, and $a_{\mathrm{res}}=\alpha_{\mathrm{res}}-\mathrm{i}\beta_{\mathrm{res}}$ is the \textit{resonant scattering length}, which sets the characteristic amplitude of the resonant variation. The parameter $\Gamma_B^{\mathrm{inel}}$ is the resonance width expressed in magnetic-field units and is associated with the finite lifetime of the quasi-bound state of the Feshbach molecule\cite{Hutson2007NewJournalofPhysics,Frye2020Phys.Rev.Research}.

In this work, we fit Eqs.~\eqref{eq:a_elastic} and \eqref{eq:a_inelastic} to the numerically calculated scattering lengths (Sec.~\ref{subsection: Coupled-channel calculations}) to extract the resonance parameters $(B_0,\Delta,a_{\mathrm{bg}})$ and, when applicable, $(a_{\mathrm{res}},\Gamma_B^{\mathrm{inel}})$ used to catalog Feshbach resonances (Sec.~\ref{subsection: 6-7 Resonances}). While overlapping-resonance formulas are available~\cite{Jachymski2013Phys.Rev.A}, we use the standard single-resonance parametrization, which is widely adopted and adequate for the resonances considered here.

Each resonance can be characterized by a dimensionless resonance strength (or pole strength) parameter $s_{\rm res}$~\cite{Chin2010Rev.Mod.Phys.}, which quantifies the coupling between the entrance channel and the resonant closed-channel bound state in an effective two-channel model near the pole: 
\begin{equation}
    s_{\rm res}=\left(\frac{a_{\rm bg}}{\bar{a}}\right)\left(\frac{\delta\mu\,\Delta}{\bar{E}}\right),
\end{equation}
where $a_{\rm bg}$ and $\Delta$ are defined in Eq. \ref{eq:a_elastic}, and $\delta\mu$ is the difference in magnetic moments between the entrance-channel threshold and the resonant bound state, evaluated in a range of $B$ where $\delta\mu$ is approximately constant. Here, $\bar{a}$ and $\bar{E}$ are the characteristic van der Waals length and energy scales,
$
\bar{a}=\frac{2\pi}{\Gamma\!\left(\tfrac14\right)^2}\left(\frac{2\mu C_6}{\hbar^2}\right)^{1/4},
$ and $
\bar{E}=\frac{\hbar^2}{2\mu \bar{a}^2},
$, where $\mu$ is the two-body reduced mass and $C_6$ is the van der Waals coefficient. Using $C_6=1393.39\,E_{\mathrm{h}}a_0^6$, $a_0$ being the Bohr radius and $E_{\mathrm{h}}$ being the Hartree energy, we obtain $\bar{a}=29.884\,a_0$ and $\bar{E}/h=671.93~\mathrm{MHz}$ for ${}^{6}$Li--${}^{6}$Li,  $\bar{a}=31.056\,a_0$ and $\bar{E}/h=533.41~\mathrm{MHz}$ for ${}^{7}$Li--${}^{7}$Li, and $\bar{a}=30.442\,a_0$ and $\bar{E}/h=601.336~\mathrm{MHz}$ for ${}^{6}$Li--${}^{7}$Li. In practice, we extract \(\delta\mu=\partial(E_{\rm at}-E_{\rm mol})/\partial B\) from coupled-channel calculations, where $E_{\rm at}$ is the entrance-channel threshold energy [solid curves in Fig.~\ref{fig:zeeman_shifts}(a)] and $E_{\rm mol}$ is the energy of the corresponding molecular bound state leading to resonance [dotted curves in Fig.~\ref{fig:zeeman_shifts}(a)]. For the 543- and 832-G resonances of ${}^{6}$Li--${}^{6}$Li we find $\delta\mu\simeq 2\,\mu_{\rm B}$~\cite{Chin2010Rev.Mod.Phys.} (i.e., $\delta\mu/h\simeq 2.8~\mathrm{MHz/G}$), and for the 737-G resonance of ${}^{7}$Li--${}^{7}$Li we find $\delta\mu\simeq 1.93\,\mu_{\rm B}$~\cite{Chin2010Rev.Mod.Phys.}.

For some resonances, extracting the differential magnetic moment \(\delta\mu=\partial(E_{\rm at}-E_{\rm mol})/\partial B\) is not straightforward. In such cases, we use the product $a_{\rm bg}\Delta$---or, equivalently for decaying resonances, $-\tfrac{1}{2}a_{\rm res}\Gamma$---as an alternative measure for the relative pole strength\cite{Hutson2007NewJournalofPhysics}.

\subsection{Theoretical model}

In this section we introduce the coupled channel framework used to describe ultracold Li–Li collisions and to predict scattering and bound-state observables near the threshold.

\subsubsection{Hamiltonian and interaction potentials}

The interaction Hamiltonian for two alkali atoms in their electronic ground state \(({}^{2}\!S_{1/2})\) is\cite{Berninger2013Phys.Rev.A, Julienne2014Phys.Rev.A}
\begin{equation}
\label{eq:hamiltonian}
\hat H
=\frac{\hbar^{2}}{2\mu}\!\left[-\,R^{-1}\frac{d^{2}}{dR^{2}}R+\frac{\hat L^{2}}{R^{2}}\right]
+\hat h_a+\hat h_b+\hat V(R),
\end{equation}
where \(\mu\) is the reduced mass of the two atoms, \(R\) the internuclear separation, and \(\hat L\) the relative orbital angular momentum. The single-atom Hamiltonian for atom \(j\in\{a,b\}\) includes hyperfine and Zeeman terms,
\begin{equation}
\label{eq:singleH}
\hat h_j=\zeta_j\,\hat{\boldsymbol{\imath}}_j\!\cdot\!\hat{\boldsymbol{s}}_j
+ g_e\,\mu_{\mathrm B}\,B\,\hat s_{zj}
+ g_{i,j}\,\mu_{\mathrm N}\,B\,\hat \imath_{zj},
\end{equation}
with electron (nuclear) spin operators \(\hat{\boldsymbol{s}}_j\) \((\hat{\boldsymbol{\imath}}_j)\), hyperfine constant \(\zeta_j\), electron \(g\) factor \(g_e\), nuclear \(g\) factor \(g_{i,j}\), and Bohr (nuclear) magneton \(\mu_{\mathrm B}\) \((\mu_{\mathrm N})\).

The interaction potential operator $\hat{V}(R)$ is decomposed into singlet $\left(S=0, X^1 \Sigma^{+}\right)$and triplet $(S=1$, $\left.a^3 \Sigma^{+}\right)$ adiabatic Born-Oppenheimer (ABO) electronic  potentials $V_{0(1)}(R)$  plus weak spin-spin couplings~\cite{Stoof1988Phys.Rev.B},
\begin{equation}
\label{eq:potdecomp}
\hat V(R)=V_0(R)\,\hat P_0+V_1(R)\,\hat P_1+\hat V^{\mathrm{d}}(R),
\end{equation}
where $\hat P_{0(1)}$ projects the total interaction potential onto the singlet (triplet) $S=0\,(1)$ electronic-spin sector. These ABO potentials exclude spin-dependent hyperfine and Zeeman interactions and therefore approach zero at dissociation, as shown in Fig.~\ref{fig: scattering PEC}. The term $\hat V^{\mathrm{d}}(R)$ represents the anisotropic magnetic dipole--dipole interaction: 
\begin{equation}
\hat{V}^{\mathrm{d}}(R)
= \frac{E_{\mathrm{h}}\alpha^{2}}{(R/a_{0})^{3}}\left[\hat{s}_{a}\!\cdot\!\hat{s}_{b}
-3(\hat{s}_{a}\!\cdot\!\mathbf{e}_{R})(\hat{s}_{b}\!\cdot\!\mathbf{e}_{R})\right],
\end{equation}
with \(\alpha\simeq 1/137\) the fine-structure constant, \(E_{\mathrm h}\) the Hartree energy, \(a_{0}\) the Bohr radius, and \(\mathbf{e}_{R}=\mathbf{R}/R\) the unit vector along the internuclear axis. For \(L = 0\) channels the diagonal matrix element of \(\hat V^{\mathrm{d}}\) vanishes, whereas for \(L\neq 0\) it is nonzero; off-diagonal matrix elements couple partial waves of the same parity with \(\Delta L=\pm 2\) (e.g., \(s\) to \(d\)). This anisotropy can drive weak spin relaxation in nominally elastic \(s\)-wave collisions, making the collisions inelastic and converting otherwise non-decaying resonances [Eq.~\ref{eq:a_elastic}] into decaying resonances [Eq.~\ref{eq:a_inelastic}].

We represent $V_0(R)$ and $V_1(R)$ with spectroscopically fitted MLR forms for the Li$_2$ singlet~\cite{Gunton2013Phys.Rev.A} and triplet~\cite{Semczuk2013Phys.Rev.A} potentials. The analytic form of the MLR potential is given as
\begin{equation}
V_{\mathrm{MLR}}(R)= D_e\!\left[1-\frac{u_{\mathrm{LR}}(R)}{u_{\mathrm{LR}}(R_e)}\, e^{-\beta(R)\cdot y_p^{R_{e}}(R)}\right]^2 ,
\label{eq:MLR-main}
\end{equation}
where $D_e$ and $R_e$ are the dissociation energy and equilibrium distance, $u_{\mathrm{LR}}(R)$ is the long range form of the potential, $\beta(R)$ is the polynomial that controls the shape of the potential from short-to-intermediate range, and $y_p^{r_{e}}(R)$ is a dimensionless number, representing a transformation of the radial variable $R$. Near \(R_e\),  Eq. \ref{eq:MLR-main} reduces to a Morse-like well, while at long range it captures the van der Waals tail, yielding a compact, smooth, and spectroscopically accurate description of the potential across the full interaction region. The parametrization also includes adiabatic and nonadiabatic beyond Bohr-Oppenheimer (BBO) correction terms. References [\cite{Semczuk2013Phys.Rev.A,Gunton2013Phys.Rev.A}] fit the singlet and triplet potentials following this analytic form to \(\sim\!2\times10^{4}\) Li\(_2\) spectroscopic transitions across all isotopologs, reproducing bound-state levels with an average accuracy of order \(0.001~\mathrm{cm}^{-1}\)\cite{LeRoy2009TheJournalofChemicalPhysics,Dattani2011JournalofMolecularSpectroscopy}.

Although spectroscopically precise, the raw MLR potentials—determined mostly by transitions from deeply bound levels—do not by themselves ensure threshold accuracy. The last bound states that set scattering lengths and Feshbach resonance locations are highly sensitive to the short-range phases of the wave function\cite{ Gribakin1993Phys.Rev.A,Gao2000Phys.Rev.A}, and thus to the short-range details of the singlet and triplet ABO potentials. Following prior work on $\mathrm{Cs}_2$ and $\mathrm{Li}_2$~\cite{Berninger2013Phys.Rev.A,Julienne2014Phys.Rev.A}, we modify the ABO potentials by adding a small inner-wall adjustment that tunes the last bound state energies while leaving the long-range part of the potentials unchanged. Specifically, as in Ref.~\cite{Berninger2013Phys.Rev.A, Julienne2014Phys.Rev.A}, we add a quadratic \textit{shift term} for $R<R_{e,S}$,
\begin{equation}
\label{eq:shift}
V_{\mathrm{shift},S}^{(a,b)}(R)=\mathcal{S}_{S}^{(a,b)}\,(R-R_{e,S})^{2},
\end{equation}
where $(a,b)=(6,6),(7,7)~\text{or}~(6,7)$ labels the isotopolog, $S$ labels the ABO potential, and $\mathcal{S}$ represents the shift parameter. The equilibrium bond length $R_{e,S}$ are taken from the fitted MLR potentials and are common to all isotopologs ($R_{e,0} = 2.6729874$ \AA~\cite{Gunton2013Phys.Rev.A}, $R_{e,1} = 4.170006$ \AA~\cite{Semczuk2013Phys.Rev.A}). This modification allows adjustment of the last bound levels without altering the total number of bound states. 

In Sec.~\ref{subsection: Fitting and Validation of the Potentials}, the shift parameters $\mathcal{S}_{0}^{(6,6)},\,\mathcal{S}_{1}^{(6,6)},\,\mathcal{S}_{0}^{(7,7)}$, and $\mathcal{S}_{1}^{(7,7)}$ enter as free parameters in the fit to the ${}^{6}\mathrm{Li}$--${}^{6}\mathrm{Li}$ and ${}^{7}\mathrm{Li}$--${}^{7}\mathrm{Li}$ threshold data, while the original MLR potentials are held fixed. This yields markedly improved agreement for threshold observables with the experiments over Ref.~\cite{Julienne2014Phys.Rev.A}. Inevitably, the inner-wall perturbation shifts the deeply bound levels. We discuss this drawback further in Sec.~\ref{sec:Discussions}.

\subsubsection{Coupled-channel calculations} \label{subsection: Coupled-channel calculations}

We numerically solve the Schr\"odinger equation
\begin{equation}
\label{Eq:Schrodinger}
    \hat{H}\,\Psi(R)=E\,\Psi(R)
\end{equation}
using the coupled-channel method \cite{Hutson2019ComputerPhysicsCommunications}. In this formulation, the total radial wave function is expanded in a channel basis set as
\begin{equation}
\label{Eq:totalWave}
    \Psi(R)=\frac{1}{R}\sum_{j}\psi_{j}(R)\,\ket{j(\tau)},
\end{equation}
where $\psi_{j}(R)$ and $\ket{j(\tau)}$ are the radial wave function and the channel function of the $j$th channel, respectively. $\tau$ represents all degrees of freedom other than $R$. 

Substituting Eq.~\eqref{Eq:totalWave} into Eq.~\eqref{Eq:Schrodinger} yields a set of coupled differential equations,
\begin{equation}
\label{Eq:CC}
    \left[-\frac{\hbar^{2}}{2\mu}\frac{\mathrm{d}^{2}}{\mathrm{d}R^{2}}\right]\psi_{j}(R)
    +\sum_{k} W_{jk}(R)\,\psi_{k}(R)
    =E\,\psi_{j}(R),
\end{equation}
where the basis-dependent \textit{coupling matrix} elements write
\begin{equation}
\label{Eq:CouplingConst}
    W_{jk}(R)=\bra{j}\left(\hat{h}_{a}+\hat{h}_{b}+\hat{V}(R)+\frac{\hat{L}^{2}}{2\mu R^{2}}\right)\ket{k}.
\end{equation}
In the atomic (AZB) basis---the eigenbasis of the asymptotic atomic Hamiltonian [Eq.~\ref{eq:singleH}] and the basis in which the laboratory-frame $S$ matrix is typically defined---the diagonal elements $W_{jj}(R)$ represent the separated-atom channel energies in the $R\!\to\!\infty$ limit (solid horizontal lines in Fig.~\ref{fig: scattering PEC}). The off-diagonal elements $W_{jk}(R)$ with $j\neq k$ describe the interchannel couplings.

Using the frame transformations described in Appendix~\ref{sec:frame}, the channel functions $\ket{j(\tau)}$---and hence $W_{jk}(R)$---can be expressed in an alternative basis, such as a MB basis labeled by approximate molecular quantum numbers, which enables the analysis of the spin character of the Feshbach molecules in Sec.~\ref{sec:spincharacter}.

Following standard practice for alkali--alkali scattering, we perform coupled channel calculations using the \textsc{molscat}, \textsc{bound}, and \textsc{field} packages\cite{Hutson2019ComputerPhysicsCommunications,Hutson2019ComputerPhysicsCommunicationsa}. Briefly, Eq.~\eqref{Eq:CC} is solved by propagating the radial wave functions and matching them to the appropriate boundary conditions. The bound-state energies and $S$-matrix elements are then extracted from the radial solutions. The basis set used in our calculations includes all relevant spin channels with the same $M_F$ and partial waves up to $d$ wave ($L=2$). The $S$ matrix is then used to compute the magnetic-field-dependent scattering length, from which the $B$ values of the poles and zeros in the scattering length are extracted. The calculated bound-state energies and pole- or zero positions are compared to their measured counterparts in a least-squares fitting of our potential model, as described in Sec.~\ref{Results}.

\section{Results} \label{Results}

\subsection{Fitting and validation of the potentials} \label{subsection: Fitting and Validation of the Potentials}

We perform a weighted least-squares fit of the four modified MLR interaction potentials,
\begin{equation} \label{eq: modifiedMLR}
    V_S^{(a,b)} = V_{\text{MLR},S}^{(a,b)} + V_{\text{shift},S}^{(a,b)} \quad [S = 0, 1; (a,b) = (6,6),(7,7)]
\end{equation}
to the measured bound-state and scattering observables of the
$^{6}$Li--$^{6}$Li and $^{7}$Li--$^{7}$Li. The only adjustable parameters are the
shift parameters entering $V_{\mathrm{shift},S}^{(A)}$ [Eq.~\eqref{eq:shift}], while the underlying
MLR potentials $V_{\mathrm{MLR},S}^{(A)}$ are held fixed. The experimental data set is identical to
that used in Ref.~\cite{Julienne2014Phys.Rev.A}, comprising precisely determined bound-state energies
$E_b$ (reported as frequencies $\nu_b=E_b/h$) \cite{Bartenstein2005Phys.Rev.Lett.,Zuern2013Phys.Rev.Lett.,Dyke2013Phys.Rev.A}
and Feshbach-resonance pole- or zero locations (reported as $B$ field values)\cite{Du2008Phys.Rev.Lett.,Hazlett2012Phys.Rev.Lett.,Pollack2009Phys.Rev.Lett.,Gross2011ComptesRendusPhysique}.
Results of our fit are summarized in Table~\ref{tab:mlr-params}, alongside the corresponding results from Ref.~\cite{Julienne2014Phys.Rev.A}.

\begin{table*}[tbp]
	\caption{
    Fitting the Li--Li interaction potentials to the measured bound-state and scattering observables of the $^{6}$Li--$^{6}$Li and $^{7}$Li--$^{7}$Li isotopologs. The ``Quantity" and ``Experiment" columns list the observables used for the fit and their measured values (with quoted uncertainties in parentheses), respectively. The rf transition frequencies ($\nu_{b,ij}$) are reported in units of kilohertz, while the locations of the poles and zeros are in units of G. The fit results are presented as the optimized shift parameters ($\mathcal{S}_{0}^{(6,6)},\,\mathcal{S}_{1}^{(6,6)},\,\mathcal{S}_{0}^{(7,7)}$, and $\mathcal{S}_{1}^{(7,7)}$) and predicted observable values (this work). The optimized shift parameters are reported in units of $10^{-6}\,E_{\mathrm{h}}/a_0^2$. 95\% confidence intervals for these values are shown in parentheses. The fit quality is quantified using the chi-squared ($\chi^2$) and reduced chi-squared ($\chi_\nu^2$) statistics, as defined in the main text; we also present the per-observable chi-squared contributions ($\chi_i^2$). For comparison, we show the fit results from Ref.~\cite{Julienne2014Phys.Rev.A} alongside ours; the corresponding $\chi^2$ values are computed by us from the results reported there.} 

\label{tab:mlr-params}
\centering
\scriptsize
\setlength{\extrarowheight}{-1pt}   % tighten row padding
\renewcommand{\arraystretch}{0.85}  % compress row height (lower = tighter)
\begin{tabular}{c c c c c c}
\toprule\toprule
Quantity & {Experiment} & {Ref.~\cite{Julienne2014Phys.Rev.A}} & $\chi^2_i$ & {This work} & $\chi^2_i$ \\
\midrule
\multicolumn{6}{c}{$^6\mathrm{Li}$--${}^{6}\mathrm{Li}$} \\
$\nu_{b,12}-\nu_{b,13}+\nu_{ff}$ at 661.436~G & 83664.5(10)\cite{Bartenstein2005Phys.Rev.Lett.} & 83665.9(8) & 1.960 & 83665.9(1) & 1.960 \\
$\nu_{b,12}-\nu_{b,13}+\nu_{ff}$ at 676.090~G & 83296.6(10)\cite{Bartenstein2005Phys.Rev.Lett.} & 83297.3(5) & 0.490 & 83297.3(1) & 0.490 \\
$\nu_{b,12}$ at 720.965~G & 127.115(31)\cite{Zuern2013Phys.Rev.Lett.} & 127.115(58) & 0.000 & 127.14(6) & 0.650 \\
$\nu_{b,12}$ at 781.057~G & 14.157(24)\cite{Zuern2013Phys.Rev.Lett.} & 14.103(37) & 5.063 & 14.10(2) & 5.641 \\
$\nu_{b,12}$ at 801.115~G & 4.341(50\cite{Zuern2013Phys.Rev.Lett.}) & 4.342(24) & 0.000 & 4.34(1) & 0.000 \\
$v_{b,12}$ at 811.139~G & 1.803(25)\cite{Zuern2013Phys.Rev.Lett.} & 1.828(16) & 1.000 & 1.824(7) & 0.706 \\
Zero in $a_{12}$ & 527.5(2)\cite{Du2008Phys.Rev.Lett.} & 527.32(8) & 0.810 & 527.175(5) & 2.641 \\
Narrow pole in $a_{12}$ & 543.286(3)\cite{Hazlett2012Phys.Rev.Lett.} & 543.41(12) & 1708.444 & 543.286(5) & 0.000 \\
\midrule
\multicolumn{2}{l}{Total $\chi^2$ ($^6\mathrm{Li}$--${}^{6}\mathrm{Li}$)} & & 1717.8 & & 12.1 \\
\multicolumn{2}{l}{Reduced $\chi^2_{\rm red}$ ($^6\mathrm{Li}$--${}^{6}\mathrm{Li}$, $N_{\rm obs}-2=6$)} & & 286.29 & & 2.01 \\
\addlinespace
\multicolumn{6}{l}{} \\
% $a_s$ ($a_0$) &  & 45.154(2) & & 45.174(3) & \\
% $a_t$ ($a_0$) &  & -2113(2) & & -2110(2) & \\
$\mathcal{S}_{0}^{(6,6)}$ ($10^{-6}\,E_{\mathrm{h}}/a_0^2$) &  & -11.8959(383) & & 16.031(5) & \\
$\mathcal{S}_{1}^{(6,6)}$ ($10^{-6}\,E_{\mathrm{h}}/a_0^2$) &  &0.51031(203) & & 1.6987(9) & \\
\addlinespace
\multicolumn{6}{c}{$^7\mathrm{Li}$--${}^{7}\mathrm{Li}$} \\
Zero in $a_{11}$ & 543.6(1)\cite{Pollack2009Phys.Rev.Lett.} & 543.64(19) & 0.160 & 543.6(1) & 0.000 \\
Pole in $a_{22}$ & 844.9(8)\cite{Gross2011ComptesRendusPhysique} & 845.31(4) & 0.263 & 845.12(3) & 0.076 \\
Pole in $a_{22}$ & 893.7(4)\cite{Gross2011ComptesRendusPhysique}  & 893.78(4) & 0.040 & 894.05(2) & 0.766 \\
$\nu_{b,11}$ at 736.8~G & 40(3)\cite{Dyke2013Phys.Rev.A} & 34.3(1.0) & 3.610 & 36(1) & 1.778 \\
$\nu_{b,11}$ at 736.5~G & 62(2)\cite{Dyke2013Phys.Rev.A} & 61.4(1.3) & 0.090 & 63(1) & 0.250 \\
$\nu_{b,11}$ at 735.5~G & 212(2)\cite{Dyke2013Phys.Rev.A} & 209.2(2.3) & 1.960 & 211(3) & 0.250 \\
$\nu_{b,11}$ at 734.4~G & 469(3)\cite{Dyke2013Phys.Rev.A} & 474.6(3.5) & 3.484 & 474(4) & 2.778 \\
$v_{b,11}$ at 733.5~G & 775(9)\cite{Dyke2013Phys.Rev.A} & 772(5) & 0.111 & 768(5) & 0.605 \\
$\nu_{b,11}$ at 732.1~G & 1375(10)\cite{Dyke2013Phys.Rev.A} & 1378(7) & 0.090 & 1366(8) & 0.810 \\
$\nu_{b,11}$ at 728.0~G & 4019(90)\cite{Dyke2013Phys.Rev.A} & 4114(14) & 1.114 & 4066(15) & 0.273 \\
\midrule
\multicolumn{2}{l}{Total $\chi^2$ ($^7\mathrm{Li}_2$)} & & 10.92 & & 7.58 \\
\multicolumn{2}{l}{Reduced $\chi^2_{\rm red}$ ($^7\mathrm{Li}$--${}^{7}\mathrm{Li}$, $N_{\rm obs}-2=8$)} & & 1.37 & & 0.95 \\
\addlinespace
\multicolumn{6}{l}{} \\
% $a_s$ ($a_0$) &  & 34.331(2) & & 34.360(3) & \\
% $a_t$ ($a_0$) &  & -26.92(7) & & -27.195(8) & \\
$\mathcal{S}_{0}^{(7,7)}$ ($10^{-6}\,E_{\mathrm{h}}/a_0^2$)  &  &-8.8530(810)& & 16.80(4) & \\
$\mathcal{S}_{1}^{(7,7)}$ ($10^{-6}\,E_{\mathrm{h}}/a_0^2$)  &  &1.210(152) & & 1.47(9) & \\
\midrule
\multicolumn{2}{l}{\textbf{Total $\chi^2$ (all observables, $N_{\rm obs}=18$)}} & & \textbf{1728.7} & & \textbf{19.7} \\
\multicolumn{2}{l}{\textbf{Overall reduced $\chi^2_{\rm red}$ (all observables, $N_{\rm obs}-4=14$)}} & & \textbf{123.48} & & \textbf{1.41} \\
\bottomrule\bottomrule
\end{tabular}
\end{table*}

\begin{table}[tbp]
	\caption{
Singlet ($a_s$) and triplet ($a_t$) $s$-wave scattering lengths for Li--Li collisions obtained from this work and Ref. \cite{Julienne2014Phys.Rev.A}, reported in units of the Bohr radius $a_0$. Uncertainties represent 95\% confidence intervals and are determined by the uncertainties of our fitted potentials.
}
\label{tab:a_s and a_t}
\begin{tabular}{ccccc}
    \toprule
    & \multicolumn{2}{c}{Ref. \cite{Julienne2014Phys.Rev.A}} & \multicolumn{2}{c}{This work} \\
    \cmidrule(lr){2-3} % Partial horizontal line spanning columns 2 and 3
    \cmidrule(lr){4-5}
    isotopolog & $a_s$ & $a_t$ & $a_s$ & $a_t$ \\
    \midrule
    $^{6}$Li--$^{6}$Li & 45.154(2) & -2113(2) & 45.174(3) & -2110(2) \\
    $^{7}$Li--$^{7}$Li & 34.331(2) & -26.92(7) & 34.360(3) & -27.195(8) \\
    $^{6}$Li--$^{7}$Li & - & - & -16.759(33) & 40.872(14) \\
    \bottomrule
\end{tabular}
\end{table}

We quantify the quality of the present and previous fits using the chi-squared statistic
$\chi^{2}=\sum_{i=1}^{N}\left(\frac{y_i^{\mathrm{(obs)}}-y_i^{\mathrm{(calc)}}}{\sigma_i}\right)^{2}$ and its reduced form $
\chi_{\nu}^{2}=\frac{\chi^{2}}{\nu}$, $\nu = N-M$, 
where $N$ is the number of data points, $\sigma_i$ is the uncertainty of point $i$, $M$ is the number of fitted parameters, and $\nu$ is the number of degrees of freedom. The contribution to $\chi^2$ from the $i$th observable is $\left(\frac{y_i^{\mathrm{(obs)}}-y_i^{\mathrm{(calc)}}}{\sigma_i}\right)^{2}$.
 For $^{6}$Li--$^{6}$Li, $\chi_\nu^2$ is reduced from 286, the value we calculated from the data in Ref.~\cite{Julienne2014Phys.Rev.A},  to 2.01 in the present work. The improvement is driven primarily by a more accurate prediction of the narrow ($0.1$~G) resonance at $543.286$~G in the $(1,2)$ channel.
For $^{7}$Li--$^{7}$Li, our $\chi_\nu^2=0.95$ is comparable to the previous value of 1.37.

From the fitted interaction model we also extract the singlet and triplet scattering lengths, $a_s$ and $a_t$, defined as the $s$-wave scattering lengths of the singlet and triplet ABO potentials, respectively, in the absence of spin-dependent couplings. The resulting values are reported in Table~\ref{tab:a_s and a_t} and are in reasonable agreement with those of Ref.~\cite{Julienne2014Phys.Rev.A}.

As an external validation of our fits, we compute the energies of the last bound vibrational levels of $^{6}$Li$_2$ using the optimized potentials and compare them with the spectroscopic measurements of Ref.~\cite{Semczuk2014Phys.Rev.Lett.} (Table~\ref{tab:2014_bound}). Our model reproduces the last singlet bound level $(v=38, F=0)$ to within $\sim 0.01$~MHz, and the last triplet bound levels $(v=9, F=0,1,2)$ to within 1~MHz. For these states, the agreement improves by approximately an order of magnitude compared with predictions obtained using the fitted potentials of Ref.~\cite{Julienne2014Phys.Rev.A}. We note that the measurements of Ref.~\cite{Semczuk2014Phys.Rev.Lett.} were published after Ref.~\cite{Julienne2014Phys.Rev.A} and were therefore unavailable for inclusion in their fitting or validation.

\begin{table*}[tbp]
\caption{
Comparison of measured and calculated binding energies (in units of megahertz) of the last singlet ($v = 38$) and triplet ($v = 9$) bound states of ${}^{6}\mathrm{Li}_2$. $F$ represents the quantum number for the total spin, which results from the coupling between the total electronic and nuclear spins. Calculations are performed using both the potentials developed in Ref.~\cite{Julienne2014Phys.Rev.A} and this work. $|\Delta|$ represents the absolute difference between the calculated and measured values.
}
\label{tab:2014_bound}
\centering
\begingroup
\scriptsize
\setlength{\tabcolsep}{5pt}
\begin{ruledtabular}
\begin{tabular}{@{} c c c c c c c @{}}
$v$ & $F$ & Measured (Ref.~\cite{Semczuk2014Phys.Rev.Lett.}) & Calculated (Ref.~\cite{Julienne2014Phys.Rev.A}) & $|\Delta|$ & Calculated (This work) & $|\Delta|$ \\
\hline
\multicolumn{7}{c}{$X^1\Sigma_g^{+}$} \\
38 & 2 & --- & $-1309.10(15)$ & --- & $-1308.76(2)$ & --- \\
38 & 0 & $-1321.671(21)$ & $-1321.98(12)$ & $0.31$ & $-1321.66(2)$ & $0.01$ \\
\multicolumn{7}{c}{$a^3\Sigma_u^{+}$} \\
9 & 2 & $-24010.65(5)$ & $-24019.06(25)$ & $8.41$ & $-24009.89(3)$ & $0.76$ \\
9 & 1 & $-24163.04(11)$ & $-24171.29(28)$ & $8.25$ & $-24162.12(3)$ & $0.92$ \\
9 & 0 & $-24238.37(5)$ & $-24246.99(32)$ & $8.62$ & $-24237.80(4)$ & $0.57$ \\
\end{tabular}
\end{ruledtabular}
\endgroup
\end{table*}

\subsection{\texorpdfstring{$^{6}$Li--$^{7}$Li}{6Li-7Li} resonances} \label{subsection: 6-7 Resonances}
\begin{table*}[tbp]
    \centering
    \scriptsize
    \setlength{\tabcolsep}{3pt}
\caption{
Resonance parameters for Feshbach resonances in the (1,1) channel of ${}^{6}\mathrm{Li}$--${}^{7}\mathrm{Li}$. Our predicted pole positions ($B_0$) are compared with those from prior measurements ($B_0^{\rm expt}$) and calculations ($B_0^{\rm theory}$). $(a_{\rm bg}/a_0)\Delta$ serves as an alternative characterization of the pole strength to $s_{\text{res}}$. The $214$-G resonance is too narrow to be fully characterized and is thus not listed.
}
    \begin{tabular}{@{} c c c c c c c c  @{}}
        \toprule
        Channel $(i,j)$ &
        $B_0$ (G) & $\Delta$ (mG) & $a_{\rm bg}/a_0$ & $s_{\rm res}$ &
        $B_0^{\rm expt}$~\cite{Zhang2005AIPConferenceProceedings} (G) &
        $B_0^{\rm theory}$~\cite{Kempen2004Phys.Rev.A} (G) &
        $(a_{\rm bg}/a_0)\Delta$ (mG)  \\
        \midrule
        (1,1) & 226.76(31) & 16.5214(57) & 40.852(18)  & $1.83 \times 10^{-4}$ & 226.3(6) & 230 & $674.93(38)$  \\
        (1,1) & 247.48(31) & 156.67(29)  & 40.570(14)  & $1.65 \times 10^{-3}$ & 246.0(8) & 251 & $6356(12)$  \\
        (1,1) & 544.23(57) & 13.562(46)  & 40.96(11)   & $7.92 \times 10^{-5}$ & 539.9(8) & 551 & $555.5(24)$ \\
        (1,1) & 552.43(57) & 15.952(39)  & 40.814(40)  & $9.56 \times 10^{-5}$ & 548.6(9) & 559 & $651.1(17)$ \\
        \bottomrule
    \end{tabular}
	\label{tab:Li6Li7_channel11}
\end{table*}

We use the fitted homonuclear potentials to construct interaction potentials to study the scattering properties of \(^{6}\mathrm{Li}\text{--}^{7}\mathrm{Li}\). Because the fitted shift parameters show only weak isotopolog dependence, we estimate the mixed-isotope shift as \(\mathcal{S}_{S}^{(6,7)}=\big(\mathcal{S}_{S}^{(6,6)}+\mathcal{S}_{S}^{(7,7)}\big)/2\) and take \(\big|\mathcal{S}_{S}^{(6,6)}-\mathcal{S}_{S}^{(7,7)}\big|\) as its uncertainty. Since \(\big|\mathcal{S}_{S}^{(6,6)}-\mathcal{S}_{S}^{(7,7)}\big|\) greatly exceeds the fitted uncertainties of \(\mathcal{S}_{S}^{(6,6)}\) and \(\mathcal{S}_{S}^{(7,7)}\), it dominates the error budget of the mixed-isotope resonance parameters reported below. Adding these shift terms to the MLR potentials, which already include spectroscopic BBO corrections, yields updated singlet and triplet potentials for \(^{6}\mathrm{Li}\text{--}^{7}\mathrm{Li}\). We then use these potentials in coupled-channel calculations to obtain the scattering lengths and bound-state energies.

\begin{table*}[tbp]
    \centering
    \scriptsize
    \setlength{\tabcolsep}{1.6pt}
    \caption{
Resonance parameters for Feshbach resonances in select excited entrance channels of ${}^{6}\mathrm{Li}$--${}^{7}\mathrm{Li}$. $(a_{\rm bg}/a_0)\Delta$ serves as the appropriate characterization of the relative resonance strength for decaying resonances.
}
    \begin{adjustbox}{max width=\textwidth}
        \begin{tabular}{@{} l c c c c c c @{}}
            \toprule
            Channel $(i,j)$ &
            $B_0$ (G) & $\Delta$ (mG) & $a_{\rm bg}/a_0$ & $a_{\rm res}/a_0$ & $\Gamma_B^{\rm inel}$ (mG) &
            $(a_{\rm bg}/a_0)\Delta$ (mG) \\
            \midrule
            (2,1) & 252.38(32) & 14.9713(26) & 40.798(16) & $3.677(14)\times 10^{8}$  & $-3.322(14)\times10^{-6}$ & $610.80(26)$ \\
            (2,1) & 272.65(32) & 114.81(17)  & 40.594(14) & $1.2280(10)\times 10^{9}$ & $-7.591(20)\times10^{-6}$ & $4660.6(71)$ \\
            (2,1) & 301.59(31) & 2.8030(84)  & 40.510(22) & $6.222(16)\times 10^{7}$  & $-3.6498(43)\times10^{-6}$ & $113.55(35)$ \\
            (2,1) & 563.57(58) & 0.2650(12)  & 41.12(53)  & $7.727(37)\times10^{5}$   & $-2.820(30)\times10^{-5}$ & $10.90(15)$ \\
            (2,1) & 571.72(57) & 34.027(92)  & 40.997(31) & $1.8815(38)\times10^{8}$  & $-1.4829(20)\times10^{-5}$ & $1395.0(39)$ \\
            (2,1) & 581.21(56) & 26.691(68)  & 40.741(15) & $7.534(25)\times10^{9}$   & $-2.8867(14)\times10^{-7}$ & $1087.4(28)$ \\
            (3,1) & 301.81(32) & 58.454(44)  & 40.684(15) & $2.5740(72)\times10^{8}$  & $-1.8478(70)\times10^{-5}$ & $2378.1(20)$ \\
            (3,1) & 602.30(57) & 60.45(14)   & 40.882(15) & $2.6247(39)\times10^{8}$  & $-1.8831(28)\times10^{-5}$ & $2471.3(58)$ \\
            (3,2) & 370.22(32) & 141.40(23)  & 40.553(14) & $2.6855(19)\times10^{8}$  & $-4.2705(59)\times10^{-5}$ & $5734.2(95)$ \\
            (3,2) & 653.81(59) & 5.096(25)   & 40.874(91) & $1.4250(89)\times10^{6}$  & $-2.9235(53)\times10^{-4}$ & $208.3(11)$ \\
            (3,3) & 720.26(60) & 61.22(13)   & 40.88(12)  & $9.128(29)\times10^{6}$   & $-5.484(16)\times10^{-4}$ & $2502.7(91)$ \\
            (6,1) & 481.49(59) & 325.75(43)  & 34.551(21) & $7.504(22)\times10^{2}$   & $-2.9996(67)\times10^{1}$ & $1.1255(16) \times 10^{4}$ \\
            (6,1) & 524.80(59) & 8974.4(70)  & 28.282(21) & $1.6803(38)\times10^{4}$  & $-3.0212(66)\times10^{1}$ & $2.5381(27) \times 10^{5}$ \\
            (1,2) & 283.50(33) & 25.079(10)  & 40.792(15) & $2.3008(15)\times10^{1}$  & $-8.893(12)\times10^{1}$ & $1023.02(55)$ \\
            (1,2) & 313.49(32) & 300.11(68)  & 40.469(14) & $1.10562(50)\times10^{4}$ & $-2.1970(67)$ & $1.2145(28) \times 10^{4}$ \\
            (1,2) & 597.07(59) & 14.961(18)  & 40.90(26)  & $5.807(23)\times10^{3}$   & $-2.108(18)\times10^{-1}$ & $611.9(40)$ \\
            (1,2) & 604.90(58) & 5.583(16)   & 40.800(32) & $1.53920(60)\times10^{1}$ & $-2.9596(92)\times10^{1}$ & $227.79(68)$ \\
            (1,3) & 362.48(33) & 43.281(65)  & 40.386(14) & $1.52795(75)\times10^{1}$ & $-2.2880(52)\times10^{2}$ & $1747.9(27)$ \\
            (1,3) & 672.61(59) & 2.634(22)   & 39.25(31)  & $1.753(19)\times10^{2}$   & $-1.1795(87)$ & $103.4(12)$ \\
            (2,2) & 310.36(33) & 27.632(21)  & 40.720(15) & $5.3469(46)\times10^{1}$  & $-4.2086(15)\times10^{1}$ & $1125.18(95)$ \\
            (2,2) & 340.28(32) & 229.59(47)  & 40.479(13) & $1.47702(50)\times10^{4}$ & $-1.2584(27)$ & $9294(19)$ \\
            (2,2) & 633.31(58) & 9.732(27)   & 40.877(19) & $1.51214(85)\times10^{1}$ & $-5.261(14)\times10^{1}$ & $397.8(11)$ \\
            (2,3) & 390.53(33) & 44.903(76)  & 40.393(14) & $2.59863(35)\times10^{1}$ & $-1.3959(27)\times10^{2}$ & $1813.8(31)$ \\
            (2,3) & 692.37(60) & 129.21(29)  & 40.910(23) & $5.483(34)\times10^{5}$   & $-1.928(17)\times10^{-2}$ & $5286(12)$ \\
            (2,3) & 699.85(59) & 2.849(14)   & 40.19(12)  & $5.6651(29)\times10^{1}$  & $-4.042(21)$ & $114.50(66)$ \\
            \bottomrule
        \end{tabular}
    \end{adjustbox}
    \label{tab:Li6Li7_inelastic_channels}
\end{table*}

\begin{figure}[tbp]
    \centering
    \includegraphics[width=1.0\columnwidth]{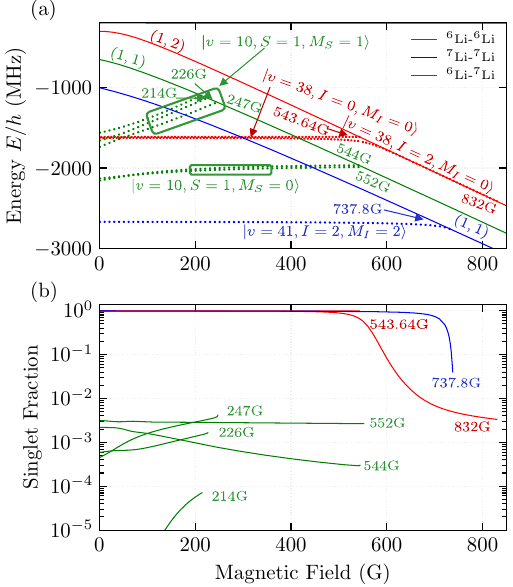}
    \caption{
    (a) Zeeman shifts of the lowest hyperfine thresholds (solid curves) and $s$-wave molecular bound states (dotted curves) for ${}^{6}$Li--${}^{6}$Li (red), ${}^{7}$Li--${}^{7}$Li (blue), and ${}^{6}$Li--${}^{7}$Li (green). The bound-state curves are labeled by appropriate vibrational and spin quantum numbers. Crossing points between the bound states and the hyperfine thresholds correspond to Feshbach resonances and are indicated by arrows. (b) Singlet fractions of the molecular bound states as a function of $B$.
    }
    \label{fig:zeeman_shifts}
\end{figure}
The lowest-energy hyperfine channel $(1,1)$ is of particular interest because its associated resonances are nondecaying, making them especially useful for tuning interaction strengths and for creating Feshbach molecules in a $^{6}$Li--$^{7}$Li mixture. From the calculated $s$-wave scattering length over $B=0$--$1000~\mathrm{G}$, we identify five poles at $214.5$, $226.4$, $247.2$, $543.6$, and $551.9~\mathrm{G}$, with widths $\Delta = 0.008$, $16.5$, $157$, $13.6$, and $16.0~\mathrm{mG}$, respectively, and a common background scattering length $a_{\mathrm{bg}}\approx 41\,a_0$. These resonances are orders of magnitude narrower than the hundreds-of-gauss-wide resonances associated with the lowest hyperfine channels of the homonuclear systems~\cite{Chin2010Rev.Mod.Phys., Julienne2014Phys.Rev.A}.

The bound states responsible for the $(1,1)$ resonances are hyperfine levels within the $(a^3\Sigma^+,\, v=10,\, N=0)$ rovibronic manifold. Their energies are plotted as a function of $B$ in Fig.~\ref{fig:zeeman_shifts}(a), together with the $(1,1)$-channel threshold. Near threshold, these bound states split into a triplet with $M_S=1$ and a doublet with $M_S=0$. The resonance positions indicated by the field values at which the bound-state curves intersect the threshold are in agreement with the pole locations extracted from the scattering length.

The five $(1,1)$ resonances were previously predicted by Kempen \textit{et~al.}~\cite{Kempen2004Phys.Rev.A} and, with the exception of the extremely narrow 214-G resonance, were experimentally observed by Zhang \textit{et~al.}~\cite{Zhang2005AIPConferenceProceedings}. We compare our results with these values in Table~\ref{tab:Li6Li7_channel11}. For the $\sim$240~G resonances, our predictions agree with the measured positions to within 1~G, consistent with the experimental uncertainties. In contrast, for the $\sim$550~G resonances, our predicted positions differ by several gauss, exceeding the experimental uncertainty. Overall, our predictions show improved agreement with experimental data compared to Ref.~\cite{Kempen2004Phys.Rev.A}, where the accuracy of their predicted resonance positions was limited in part by the use of purely mass-scaled ${ }^6 \mathrm{Li}-{ }^6 \mathrm{Li}$ interaction potentials. 
% Their potentials neglected BBO corrections, which are required for a consistent description of both scattering and spectroscopic data. 
We outline possible strategies for reconciling the remaining discrepancies with experiment in Sec.~\ref{sec:Discussions}.

We next investigate Feshbach resonances in several higher-energy hyperfine channels, all of which exhibit inelastic decay to varying degrees ($\Gamma_B^{\mathrm{inel}}>0$). The extracted resonance parameters are summarized in Table~\ref{tab:Li6Li7_inelastic_channels}. The $(6,1)$ channel features an unusually broad resonance, with width of order $\sim 10$~G, centered at $524.8$~G—orders of magnitude wider than the other resonances identified here (all of which are subgauss)—and correspondingly the largest relative pole strength, $(a_{\mathrm{bg}}/a_0)\Delta \sim 2\times 10^{5}$~mG. This is consistent with an earlier prediction of a resonance near $\sim 530$~G for this channel and with the pronounced atom-loss feature observed experimentally between $\sim 440$ and $\sim 540$~G~\cite{Zhang2005AIPConferenceProceedings}. However, using the standard relations between $\Gamma_B^{\mathrm{inel}}$ and lifetimes~\cite{Chin2010Rev.Mod.Phys.,Frye2020Phys.Rev.Research}, we estimate that the associated resonant state is extremely short lived ($\tau\sim 2~\mu\mathrm{s}$), making it unlikely to be useful for coherent applications such as molecule formation via STIRAP. Finally, the two resonances in the $(3,2)$ channel appear consistent with the resonance structure calculated in Ref.~\cite{Abeelen1997Phys.Rev.A}, although explicit resonance parameters were not reported there.

\subsection{Spin characters of Feshbach molecules}
\label{sec:spincharacter}
Next, we turn our attention to the wave functions of near-threshold bound states, often termed ``Feshbach molecules" (FMs). This analysis is motivated by the application of producing deeply bound Li$_2$ molecules from the FMs via STIRAP, a Raman process comprising an ``up-leg" transition between the FM and an electronically excited intermediate state (EM), and a ``down-leg" transition between the intermediate state and the target deeply bound state (DM). The $\langle \text{FM} | \text{EM} \rangle$ and $\langle \text{FM} | \text{DM} \rangle$ wave-function overlaps directly set the corresponding transition strengths and therefore the efficiency of the population transfer. The spin and spatial components of the FM wave function both contribute to these overlaps. Here, we focus on the spin part and defer analysis of the spatial wave functions (i.e., Franck-Condon overlaps) to a separate manuscript.

The total electron-spin quantum number $S$ of the FM is of particular interest, since the target deeply bound states generally have well-defined singlet ($S = 0$) or triplet ($S = 1$) character. We therefore express the total coupled-channel wave function of the FM in the MB as
\begin{equation}
\ket{\mathrm{FM}(R,B)}
=
\sum_{S=0,1}\ \sum_{n}
\psi^{(\mathrm{MB})}_{S, n}(R,B)\,\ket{S,n}_{\mathrm{MB}},
\label{eq:fm_mb_expansion}
\end{equation}
where $n$ represents all quantum numbers other than $S$ in the molecular basis, and the radial wave functions satisfy the normalization condition $\quad
\sum_{S,n}\int_0^\infty \left|\psi^{(\mathrm{MB})}_{S, n}(R)\right|^2\,\mathrm{d}R = 1$. The amplitudes of the radial wave functions are determined by projecting $\ket{\mathrm{FM}}$ from the TDB (the native output basis of \textsc{molscat/bound}) onto the MB, using the unitary frame transformations detailed in Appendix~\ref{sec:frame}.

We quantify the spin character of the FM using its singlet fraction, defined as
\begin{equation}
Z_{0}(B)\equiv \sum_{n}\int_0^\infty \left|\psi^{(\mathrm{MB})}_{0, n}(R)\right|^2\,\mathrm{d}R.
\label{eq:Zsing_def}
\end{equation}
The singlet fractions for FMs associated with resonances in the lowest spin channels of the three isotopologs are plotted as a function of $B$ in Fig. \ref{fig:zeeman_shifts}(b). As $B$ decreases from the pole, the FM's spin character evolves from that of the entrance channel toward that of the closed-channel bound state. For all resonances considered here, the entrance channel is primarily triplet in character, as expected for the lowest-energy hyperfine channel of alkali--alkali systems at high fields. In contrast, the closed channels are nearly pure singlet in $^6$Li$_2$ and $^7$Li$_2$, but triplet in $^6$Li$^7$Li. Thus, the homonuclear FMs exhibit a pronounced $B$ dependence in their spin character, whereas the heteronuclear FM remains predominantly triplet over the entire range considered.

The differing spin character of the least-bound states arises naturally from reduced-mass variations across isotopologs that share essentially the same electronic potentials~\cite{Flambaum1999Phys.Rev.A,Gao2000Phys.Rev.A}. Table \ref{tab:lastbound} summarizes the binding energies and quantum numbers of the least-bound vibrational levels of both the $X^1\Sigma^+$ and $a^3\Sigma^+$ potentials. In particular, the least-bound levels are $v=38$ and $v=41$ of the singlet potential for $^{6}$Li$_2$ and $^{7}$Li$_2$, respectively, versus $v=10$ of the triplet potential for $^{6}$Li$^{7}$Li.

Differences in FM spin character across isotopologs imply distinct strategies for STIRAP transfer into deeply bound molecules. In the homonuclear systems, the FM can be tuned to have predominantly singlet or triplet character by choosing an appropriate magnetic field; an intermediate state with matching spin character can then efficiently connect to deeply bound levels in either the $X^{1}\Sigma^{+}$ or $a^{3}\Sigma^{+}$ potentials. In contrast, the $^{6}$Li$^{7}$Li FM remains strongly triplet, which naturally favors transfer to deeply bound levels in the $a^{3}\Sigma^{+}$ potential. Transfer to the $X^{1}\Sigma^{+}$ potential instead requires a change in spin character and therefore an intermediate state with mixed singlet-triplet character to provide the necessary coupling\cite{Koehler2006Rev.Mod.Phys.}. Experimentally, STIRAP has so far been demonstrated only for producing triplet $^{6}$Li$_2$ molecules\cite{Polovy2020Phys.Rev.A}.

\begin{table}[htbp]
\centering
\caption{
Energies of the last-bound singlet and triplet vibrational states of Li$_2$ isotopologs (without considering hyperfine energies).
}
\label{tab:lastbound}
\begin{ruledtabular}
\begin{tabular}{lcccc}
 & \multicolumn{2}{c}{$X^1\Sigma^+$} & \multicolumn{2}{c}{$a^3\Sigma^+$} \\
\cline{2-3}\cline{4-5}
isotopolog & $v$ & Energy (GHz) & $v$ & Energy (GHz) \\
\hline
$^6$Li--$^6$Li & 38 & $-1.609301(14)$  & 9  & $-24.390475(65)$ \\
$^6$Li--$^7$Li & 39 & $-13.101(4)$     & 10 & $-1.946(2)$      \\
$^7$Li--$^7$Li & 41 & $-2.64882(18)$   & 10 & $-12.430(4)$     \\
\end{tabular}
\end{ruledtabular}
\end{table}

\subsection{Open- and closed-channel characters and resonance strengths}
\label{sec:closed_fraction}

As a complementary characterization of the Feshbach resonances, we expand $\ket{\mathrm{FM}}$ in the AZB and analyze the relative contributions of the entrance-channel scattering state and the closed-channel bound states to the resonant state. Following literature \cite{Chin2010Rev.Mod.Phys.}, we refer to the entrance channel also as the open channel in this section, since the two coincide when the entrance channel is the lowest-energy hyperfine channel. Specifically, we write
\begin{equation}
\ket{\mathrm{FM}(R,B)}
=
\sum_{i\in \mathrm{AZB}}
\psi^{(\mathrm{AZB})}_{i}(R,B)\,\ket{i}_{\mathrm{AZB}},
\label{eq:azb_expansion}
\end{equation}
where $i$ labels the AZB channels, and $
\sum_{i}\int_0^\infty \left|\psi^{(\mathrm{AZB})}_{i}(R)\right|^2\,\mathrm{d}R = 1$. The weight of the scattering wave function in $\ket{\mathrm{FM}}$ is characterized by the open channel fraction,
\begin{equation}
Z_{\mathrm{open}}(B)
\equiv
\int_0^\infty \left|\psi^{(\mathrm{AZB})}_{e}(R)\right|^2\,\mathrm{d}R,
\label{eq:Zopen_def}
\end{equation}
where $e$ labels the entrance- or open channel. Complementarily, the bound-state contribution is quantified by the closed-channel fraction $Z_{\mathrm{closed}}(B)\equiv 1-Z_{\mathrm{open}}(B)$. By definition, $Z_{\mathrm{open}}(B\to B_0) = 1$ and $Z_{\mathrm{closed}}(B\to B_0) = 0$, where $B_0$ is the pole~\cite{Chin2010Rev.Mod.Phys.}.

\begin{figure}[tbp]
  \centering
  \includegraphics[width=\columnwidth]{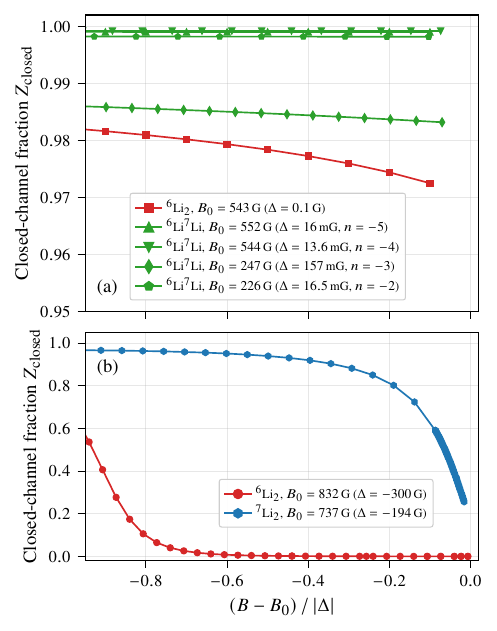}
  \caption{
  Closed-channel fraction for the lowest-energy hyperfine-channel Feshbach molecules. $(B-B_0)/|\Delta|$ represents the scaled detuning. Note that \(Z_{\mathrm{closed}}=0\) at \(B = B_0\) according to the two-channel model definition of the open- and closed-channel fractions in Ref.~\cite{Chin2010Rev.Mod.Phys.}.
  }
  \label{fig:closed_fraction}
\end{figure}
Figure ~\ref{fig:closed_fraction} shows $Z_{\mathrm{closed}}$ as a function of the scaled-field detuning $(B-B_0)/|\Delta|$ for the lowest-energy hyperfine-channel Feshbach molecules. The 832-G FM of ${}^{6}\mathrm{Li}_2$ is open-channel dominated, with $Z_{\mathrm{closed}}\approx 0$ over much of its width. By contrast, the 543-G FM of ${}^{6}\mathrm{Li}_2$ and all FMs of ${}^{6}\mathrm{Li}$${}^{7}\mathrm{Li}$ are closed-channel dominated, with $Z_{\mathrm{closed}}\approx 1$ over much of their widths. The 737-G FM of ${}^{7}\mathrm{Li}_2$ exhibits intermediate behavior, tending toward closed-channel dominance.

The behavior of the open- and closed-channel fractions as a function of $B$ is closely tied to the resonance-strength parameter $s_{\mathrm{res}}$: open-channel-dominated resonances correspond to $s_{\mathrm{res}}\gg 1$, whereas closed-channel-dominated resonances have $s_{\mathrm{res}}\ll 1$. For the homonuclear systems, $s_{\mathrm{res}}$ has previously been reported to be 59, 0.001, and 0.80 for the 832-, 543-, and 737-G resonances, respectively~\cite{Chin2010Rev.Mod.Phys.}. Here, we calculate $s_{\mathrm{res}}$ for the ${}^{6}\mathrm{Li}$--${}^{7}\mathrm{Li}$ resonances to be $10^{-5}$--$10^{-3}$, consistent with their closed-channel-dominated character (Table~\ref{tab:Li6Li7_channel11}). In evaluating $s_{\mathrm{res}}$, we use $\delta\mu/h \approx 2.69$~MHz/G for the pole near 551.874~G, $2.61$~MHz/G near 543.638~G, $4.74$~MHz/G near 247.161~G, and $4.96$~MHz/G near 226.437~G. These values of $\delta\mu/h$ are extracted from the difference between the hyperfine threshold and the relevant bound-state energies evaluated $\sim 2$--$3$~G away from each pole, in a field range where $\delta\mu$ is approximately constant.

The resonance parameters calculated here may help identify candidate magnetic-field regimes for mixed-isotope lithium few-body states, including $^{6}$Li--$^{6}$Li--$^{7}$Li and $^{7}$Li--$^{7}$Li--$^{6}$Li. These mixed-isotope states could have properties distinct from those of the familiar $^{6}$Li$_3$ and $^{7}$Li$_3$ Efimov states\cite{Braaten2006PhysicsReports,Chin2010Rev.Mod.Phys.,Greene2017Rev.Mod.Phys.,Naidon2011COMPTESRENDUSPHYSIQUE,Pollack2009Science,Gross2009Phys.Rev.Lett.,Yudkin2024NatureCommunications} when one atom is replaced by the other isotope. However, the very narrow interspecies $^{6}$Li--$^{7}$Li resonances imply limited tunability of the scattering length over most magnetic fields, restricting prospects for controlled few-body states in lithium.

\section{Discussion}
\label{sec:Discussions}

We first explain how our model differs from previous approaches. Unlike earlier work, we use spectroscopy-constrained MLR potentials that explicitly include spectroscopic BBO corrections. Reference~\cite{Kempen2004Phys.Rev.A} relied on pure mass scaling and omitted the BBO corrections required to reproduce near-threshold scattering observables across lithium isotopologs, which limited its quantitative accuracy. Although Ref.~\cite{Julienne2014Phys.Rev.A} introduced short-range shift terms, these effectively merged spectroscopic and near-threshold BBO contributions into a single empirical correction.  As a result, the shift terms obtained here exhibit smaller relative variations between isotopologs than those of Ref.~\cite{Julienne2014Phys.Rev.A} (Table~\ref{tab:mlr-params}).

\begin{figure*}[tbp]
	\centering
	\includegraphics[width=\textwidth]{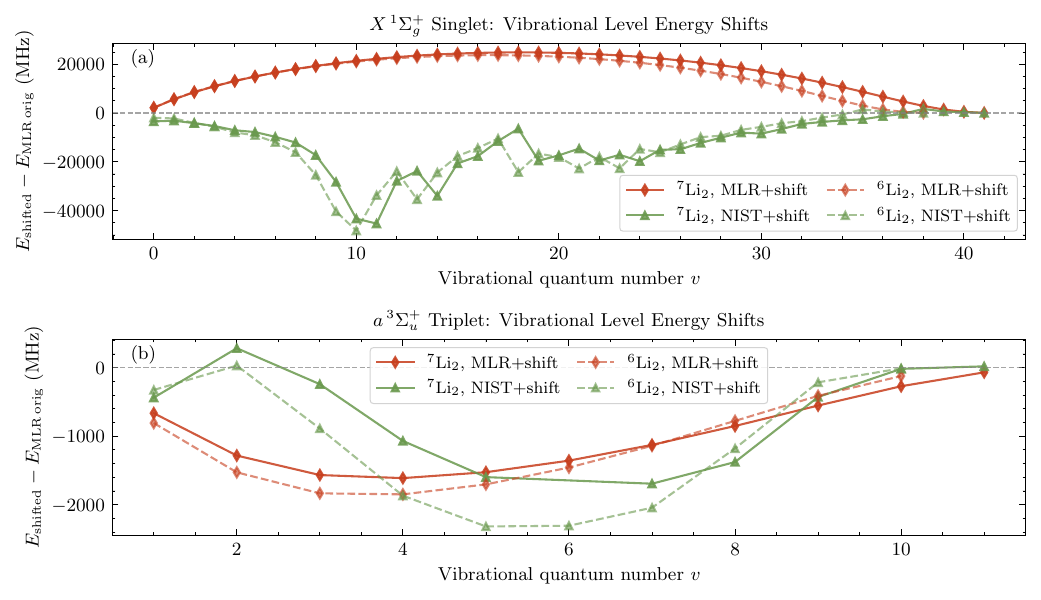}
	\caption{Vibrational-level energy shifts produced by adding the short-range shift term to the Li$_2$ interaction potentials. The ordinate shows
	\(\Delta E(v)=E_{\rm shifted}(v)-E_{\rm MLR,2013}(v)\) (MHz), referenced to the unshifted 2013 MLR potentials\cite{Gunton2013Phys.Rev.A,Semczuk2013Phys.Rev.A}; the horizontal dashed line marks \(\Delta E=0\).
	(a) Ground singlet \(X^{1}\Sigma_{g}^{+}\) manifold for $^{7}$Li$_2$ and $^{6}$Li$_2$ over the vibrational range shown.
	(b) Triplet \(a^{3}\Sigma_{u}^{+}\) manifold.
	Curves compare the shifted MLR potential optimized in this work (MLR$+$shift) and the shifted 2014/NIST potentials\cite{Julienne2014Phys.Rev.A} (NIST$+$shift). Positive (negative) \(\Delta E\) corresponds to a lower (higher) bound level relative to the 2013 MLR reference.}
	\label{fig:vibrational_shifts}
\end{figure*}

The predicted $(1,1)$-channel resonance locations from our model show improved agreement with experiment~\cite{Zhang2005AIPConferenceProceedings} relative to Ref.~\cite{Kempen2004Phys.Rev.A}, but discrepancies of order a few gauss persist (Table~\ref{tab:Li6Li7_channel11}). We attribute these residual discrepancies primarily to limitations of our interaction potentials. First, the MLR potentials are constrained mainly by homonuclear spectroscopy, whereas heteronuclear ${}^{6}\mathrm{Li}{}^{7}\mathrm{Li}$ data are comparatively sparse~\cite{LeRoy2009TheJournalofChemicalPhysics,Dattani2011JournalofMolecularSpectroscopy}. Since the observed ${}^{6}\mathrm{Li}{-}{}^{7}\mathrm{Li}$ Feshbach resonances are largely triplet-dominated,  inaccuracies in the triplet MLR potential are a plausible source of the remaining discrepancies. Second, the added shift term and our choice of its mean value are phenomenological; in particular, the arithmetic-mean prescription may not faithfully represent the threshold-specific correction required for the ${}^{6}\mathrm{Li}{-}{}^{7}\mathrm{Li}$ system.

A practical drawback of the additive shift is that it perturbs the deeply bound spectrum in a nonuniform, strongly $v$-dependent manner. Using the vibrational levels computed from the original, unshifted 2013 MLR potentials\cite{Gunton2013Phys.Rev.A,Semczuk2013Phys.Rev.A} as a reference, we define the offsets
\(\Delta E(v)=E_{\rm shifted}(v)-E_{\rm MLR,2013}(v)\),
and compare the computed vibrational levels with the shifted potentials from this work and the potential from Ref.~\cite{Julienne2014Phys.Rev.A}, which we refer to as the 2014/NIST potential. Figure~\ref{fig:vibrational_shifts} shows that the induced shifts in vibrational levels are comparatively small for the most deeply bound levels and for levels very near dissociation, but become substantially larger for intermediate vibrational levels.

To reconcile the remaining theory-experiment discrepancies for the ${}^{6}\mathrm{Li}$--${}^{7}\mathrm{Li}$ system without perturbing the deeply bound spectrum, it is desirable to refit the MLR potential simultaneously to spectroscopy data and near-threshold observables. To this end, additional spectroscopy and scattering measurements of the ${}^{6}\mathrm{Li}$--${}^{7}\mathrm{Li}$ system would be valuable.
Such a global fit would eliminate reliance on \textit{ad hoc} shift terms and yield a consistent description of both near-threshold and deeply bound states across all isotopologs.

% =================
% CONCLUSIONS
% =================
\section{Conclusions}
\label{sec:conclusions}

In this work, we refined the interaction potentials for the Li--Li system by modifying spectroscopically accurate MLR potentials with a short-range shift term and fitting the resulting potentials to threshold observables for ${}^{6}\mathrm{Li}$--${}^{6}\mathrm{Li}$ and ${}^{7}\mathrm{Li}$--${}^{7}\mathrm{Li}$. The resulting model provides an improved quantitative description of near-threshold bound states and Feshbach resonances in the homonuclear isotopologs, and yields corresponding predictions for heteronuclear ${}^{6}\mathrm{Li}$--${}^{7}\mathrm{Li}$ resonances. We further characterize the near-threshold bound states by extracting their binding energies, singlet- and triplet characters, and open- and closed-channel fractions. Together, these results provide a foundation for future studies of optical transfer (STIRAP) to deeply bound lithium molecules.

\section{Acknowledgments}
J.-C.Z. and Y.L. are supported by the startup funding from the University of Maryland, College Park. J.-C.Z. gratefully acknowledges the discussions during 2025 Quantum Control GRC and helpful comments from Zhen-Ze Li, Xing-Yan Chen, and Zeyang Li and the early discussions with Nike Dattani. 

\section{Data Availability}
There are no publicly available research data or software supporting this manuscript. Requests for further information or data should be sent to the authors.

\appendix
\section{Frame transformations}
\label{sec:frame}
%---------------------------------------

The \textsc{molscat/bound} code suite outputs the coupled-channel radial wave function in the TDB introduced in Sec.~\ref{sec:basis}~\cite{Hutson2019ComputerPhysicsCommunications,Hutson2019ComputerPhysicsCommunicationsa}. To characterize the Feshbach molecules, we expand the FM wave function, $\ket{\mathrm{FM}(R,B)}$, in two bases: the molecular basis (MB), which resolves the singlet- and triplet spin character, and the adiabatic Zeeman basis (AZB), which resolves the open- and closed-channel fractions. For convenience, we present the explicit frame transformations between the TDB, MB, and AZB used throughout this work.

\begin{figure}[tbp]
\centering
\includegraphics[width=\columnwidth]{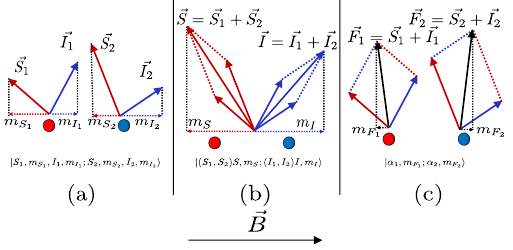}
\caption{Unitary transformations connecting the three bases used in this work, introduced in Sec.\ref{sec:basis}: 
\textbf{(a) TDB:} $\ket{m_{s,a},m_{i,a};\,m_{s,b},m_{i,b};\,\ell m_\ell}$.
\textbf{(b) MB:} $\ket{S m_S;\,I m_I;\,\ell m_\ell}$ with $\vec{S}=\vec{s}_a+\vec{s}_b$ and $\vec{I}=\vec{i}_a+\vec{i}_b$.
\textbf{(c) AZB:} $\ket{\alpha_a(B),m_{f,a};\,\alpha_b(B),m_{f,b};\,\ell m_\ell}$, constructed from the field-dependent single-atom Breit-Rabi eigenstates.}
\label{fig:transform}
\end{figure}

%---------------------------------------
\subsection{\texorpdfstring{TDB $\leftrightarrow$ MB}{TDB <-> MB} transformation}
%---------------------------------------

The transformation between the TDB and MB is field independent and is given by standard angular-momentum recoupling of electronic and nuclear spins. For fixed $\ell,m_\ell$,
\begingroup\small
\begin{align}
&\ket{S m_S;\, I m_I;\, \ell m_\ell}
\nonumber\\
&\quad=
\!\!\!\sum_{\substack{m_{s,a},m_{s,b}\\[-2pt] m_{i,a},m_{i,b}}}
\langle s_a m_{s,a}, s_b m_{s,b}\mid S m_S\rangle
\nonumber\\
&\qquad\times
\langle i_a m_{i,a}, i_b m_{i,b}\mid I m_I\rangle
\nonumber\\
&\qquad\times
\ket{m_{s,a},m_{i,a};\,m_{s,b},m_{i,b};\,\ell m_\ell}.
\label{eq:TDBtoMB}
\end{align}
\endgroup
This defines the unitary matrix $U_{\mathrm{TDB}\to\mathrm{MB}}$; its inverse is $U_{\mathrm{MB}\to\mathrm{TDB}}=U_{\mathrm{TDB}\to\mathrm{MB}}^\dagger$.

%---------------------------------------
\subsection{\texorpdfstring{TDB $\leftrightarrow$ AZB}{TDB <-> AZB} transformation}
\label{sec:breit}
%---------------------------------------

The AZB is constructed from eigenstates of the single-atom hyperfine-Zeeman Hamiltonian in a magnetic field $B\hat z$,
\begin{equation}
\hat{h}_{\rm at}(B)=\zeta\,\hat{\mathbf{i}}\!\cdot\!\hat{\mathbf{s}}
+ g_{s}\mu_{\mathrm{B}}B\, \hat{s}_z
+ g_{i}\mu_{\mathrm{N}}B\, \hat{i}_z ,
\label{eq:H_atom}
\end{equation}
where $\zeta$ is the atomic hyperfine constant, $g_s$ and $g_i$ are the electron and nuclear $g$ factors, and $\mu_B$ ($\mu_N$) is the Bohr (nuclear) magneton. The Hamiltonian is block diagonal in the conserved projection $m_f=m_s+m_i$. For $s=\tfrac12$, each block with $-i+\tfrac12 \leq m_f \leq i-\tfrac12$ is a $2\times 2$ matrix in the basis $\{\ket{\tfrac12,\,m_f-\tfrac12},\ket{-\tfrac12,\,m_f+\tfrac12}\}$, whose eigenvectors define the adiabatic (Breit-Rabi) states:
\begingroup\small
\begin{align}
\ket{\alpha{=}{+},m_f;B}
&=\cos\theta_{m_f}(B)\,\ket{\tfrac12,m_f{-}\tfrac12}
\nonumber\\
&\quad+\sin\theta_{m_f}(B)\,\ket{{-}\tfrac12,m_f{+}\tfrac12},
\nonumber\\[2pt]
\ket{\alpha{=}{-},m_f;B}
&=-\sin\theta_{m_f}(B)\,\ket{\tfrac12,m_f{-}\tfrac12}
\nonumber\\
&\quad+\cos\theta_{m_f}(B)\,\ket{{-}\tfrac12,m_f{+}\tfrac12}.
\label{eq:BRvec}
\end{align}
\endgroup
The mixing angle may be written as
\begin{equation}
\tan 2\theta_{m_f}(B)
=\frac{\zeta\sqrt{i(i{+}1)-m_f^2+\tfrac14}}
{\zeta m_f+\big(g_s\mu_{\mathrm{B}}-g_i\mu_{\mathrm{N}}\big)B}.
\label{eq:theta}
\end{equation}
States with $m_f=\pm(i+\tfrac12)$ are already eigenstates and do not mix.

Let $U^{(\mathrm{BR})}(B)$ denote the single-atom unitary matrix mapping the decoupled basis $\ket{m_s,m_i}$ to the adiabatic basis $\ket{\alpha,m_f;B}$. For two atoms $a,b$, the two-body transformation is the tensor product
\begin{equation}
U_{\mathrm{TDB}\to\mathrm{AZB}}(B)
=U^{(\mathrm{BR},a)}(B)\otimes U^{(\mathrm{BR},b)}(B),
\label{eq:TDBtoAZB}
\end{equation}
acting within each fixed $\ell,m_\ell$ block. The inverse is $U_{\mathrm{AZB}\to\mathrm{TDB}}(B)=U_{\mathrm{TDB}\to\mathrm{AZB}}^\dagger(B)$.

%---------------------------------------
\subsection{\texorpdfstring{AZB $\leftrightarrow$ MB}{AZB <-> MB} transformation}
%---------------------------------------

The AZB-MB transformation follows from composition:
\begin{align}
U_{\AZB\to\MB}(B)
&= U_{\TDB\to\MB}\;U_{\AZB\to\TDB}(B) \nonumber\\
&= U_{\TDB\to\MB}\;U_{\TDB\to\AZB}^{\dagger}(B).
\label{eq:AZBtoMB}
\end{align}

and $U_{\mathrm{MB}\to\mathrm{AZB}}(B)=U_{\mathrm{AZB}\to\mathrm{MB}}^\dagger(B)$.
%---------------------------------------
\section{Symmetries and constraints}
%---------------------------------------
\label{sec:symmetrized}
The \textsc{molscat/bound} code suite~\cite{Hutson2019ComputerPhysicsCommunications,Hutson2019ComputerPhysicsCommunicationsa} exploits conservation laws and exchange symmetry to speed up the coupled-channel calculations. In particular, calculations are carried out independently within each fixed-$M_{\rm tot}$ block, and for homonuclear dimers the appropriate symmetrization under particle exchange is enforced internally. We briefly summarize these constraints here.

\subsection{Conserved quantities}

In a magnetic field, the Hamiltonian conserves the total projection of angular momentum onto the space-fixed $z$ axis (taken along $\vec B$). In the bases of Sec.~\ref{sec:basis}, this may be written equivalently as
\begin{equation}
M_{\rm tot}
= m_{f,a}+m_{f,b}+m_\ell
= m_S+m_I+m_\ell
= M_F+m_\ell,
\label{eq:Mtot}
\end{equation}
where $M_F=m_S+m_I=m_{f,a}+m_{f,b}$. This conservation law block diagonalizes the coupled-channel problem, and all frame transformations may be performed independently within each fixed-$M_{\rm tot}$ block.

\subsection{Symmetrized basis for identical particles}

For homonuclear dimers, the total two-atom wave function must have definite symmetry under exchange of the two identical atoms.  For example, the symmetrized AZB channel functions can be written as: 
\begin{align}
\ket{\{\beta_1\beta_2\}^{\rho};\,\ell m_\ell}
&= \mathcal{N}\Big(
\ket{\beta_1}_a\ket{\beta_2}_b
+\rho(-1)^{\ell}\ket{\beta_2}_a\ket{\beta_1}_b
\Big)\nonumber\\
&\qquad\times\ket{\ell m_\ell},
\label{eq:sym_AZB}
\end{align}
where $\ket{\beta}_\kappa\equiv\ket{\alpha_\kappa(B),m_{f,\kappa}}$ denotes a single-atom AZB state with $\kappa=a,b$, and $\rho=\pm1$ is the eigenvalue under particle exchange. The normalization factor is $\mathcal{N}=1/\sqrt{2}$ when $\beta_1\neq\beta_2$. For $\beta_1=\beta_2$, the symmetrized state exists only if $\rho(-1)^{\ell}=+1$, in which case $\mathcal{N}=1/2$.

Quantum statistics fixes the allowed exchange symmetry: for two identical bosons (fermions), the internal$\times$rotational two-body basis must be symmetric (antisymmetric) under exchange, corresponding to $\rho=+1$ ($\rho=-1$). For lithium, ${}^{6}$Li is fermionic and ${}^{7}$Li is bosonic; heteronuclear ${}^{6}$Li--${}^{7}$Li collisions have no exchange-symmetry constraint. Enforcing this identical-particle symmetry reduces the size of the coupled-channel basis sets and simplifies the calculations.

% \bibliographystyle{apsrev4-2}
% \vspace*{\fill}
\input{feshbach_paper_arxiv.bbl}
% \bibliography{feshbach}

\end{document}